\documentclass[fleqn,usenatbib,useAMS]{mnras}
\let\oldhref\href
\renewcommand{\href}[2]{\oldhref{#1}{\hbox{#2}}}

\usepackage{supertabular}
\usepackage{graphicx}	
\usepackage{amsmath}	
\usepackage{amssymb}	
\usepackage{multicol}        
\usepackage{bm}		

\defcitealias{lehner11}{LH11}
\defcitealias{lehner12}{L12}
\defcitealias{lehner22}{Paper I}

\newcommand{\rAA}{\rm \AA}

\newcommand{\km}{${\rm km\,s}^{-1}$}
\newcommand {\kpc} {\,{\rm kpc}}
\newcommand{\kms} {\,{\rm km\,s}^{-1}}

\newcommand {\cmmq}{\,{\rm cm^{-2}}}
\newcommand{\hi} {{\rm H}\,{\small\rm I}}

\newcommand{\alii}{Al$\;${\small\rm II}\relax}

\newcommand{\cii}{C$\;${\small\rm II}\relax}

\newcommand{\civ}{C$\;${\small\rm IV}\relax}

\newcommand{\oi}{O$\;${\small\rm I}\relax}

\newcommand{\sii}{S$\;${\small\rm II}\relax}
\newcommand{\siii}{Si$\;${\small\rm II}\relax}
\newcommand{\siiii}{Si$\;${\small\rm III}\relax}

\newcommand{\siiv}{Si$\;${\small\rm IV}\relax}

\newcommand{\feii}{Fe$\;${\small\rm II}\relax}

\defcitealias{marasco19}{M19}

\title[HVC/IVC kinematics]{Intermediate- and high-velocity clouds in the Milky Way II: evidence for a Galactic fountain with collimated outflows and diffuse inflows}

\author[Marasco et al.]{Antonino Marasco$^1$,  Filippo Fraternali$^2$,  Nicolas Lehner$^3$, J. Christopher Howk$^3$\\
$1$ INAF - Osservatorio Astrofisico di Arcetri, Largo E. Fermi 5, 50127, Firenze, Italy \\
$2$ Kapteyn Astronomical Institute, University of Groningen, Postbus 800, 9700 AV Groningen, The Netherlands \\
$3$ Department of Physics, University of Notre Dame, Notre Dame, IN 46556, USA \\
}

\date{Last updated \today; in original form \today}

\pubyear{2022}

\begin{document}
\label{firstpage}
\pagerange{\pageref{firstpage}--\pageref{lastpage}}
\maketitle

\begin{abstract}
We model the kinematics of the high- and intermediate- velocity clouds (HVCs and IVCs) observed in absorption towards a sample of 55 Galactic halo stars with accurate distance measurements.
We employ a simple model of a thick disc whose main free parameters are the gas azimuthal, radial and vertical velocities ($v_\phi$, $v_{\rm R}$ and $v_{\rm z}$), and apply it to the data by fully accounting for the distribution of the observed features in the distance-velocity space.
We find that at least two separate components are required to reproduce the data.
A scenario where the HVCs and the IVCs are treated as distinct populations provides only a partial description of the data, which suggests that a pure velocity-based separation may give a biased vision of the gas physics at the Milky Way's disc--halo interface.
Instead, the data are best described by a combination of an inflow and an outflow components, both characterised by rotation with $v_\phi$ comparable to that of the disc and $v_{\rm z}$ of $50\!-\!100\kms$.
Features associated with the inflow appear to be diffused across the sky, while those associated with the outflow are mostly confined within a bi-cone pointing towards ($l\!=\!220^{\circ}$, $b\!=\!+40^{\circ}$) and ($l\!=\!40^{\circ}$, $b\!=\!-40^{\circ}$).
Our findings indicate that the lower ($|z|\!\lesssim\!10\kpc$) Galactic halo is populated by a mixture of diffuse inflowing gas and collimated outflowing material, which are likely manifestations of a galaxy-wide gas cycle triggered by stellar feedback, that is, the galactic fountain. 

\end{abstract}


\section{Introduction}\label{s-intro}
The evolution of late-type galaxies is largely controlled by the cycle of gas between their star forming disc and their gaseous halo.
Galaxies like the Milky Way require a continuous supply of low-metallity material to replenish the gas used to form stars \citep{chiappini01, FraternaliTomassetti12, Saintonge+13}, and most of this gas is expected to originate from smooth accretion from the halo rather than from wet mergers \citep{sancisi+08,diteodorofraternali14}.
Feedback from star formation and active galactic nuclei is thought to generate galaxy-scale outflows where the gas can, depending on the launching conditions, be expelled beyond the virial radius of the galaxy \citep{Muratov+15,Mitchell+20}, be dissolved within the low-density circumgalactic medium \citep[CGM;][]{heitsch09}, or return back to the disc in a `galactic fountain' (GF) cycle \citep{shapiro76,bregman80,fraternali06}.
Ultimately, the combinations of inflow and outflow processes regulates the main properties (mass, angular momentum, metallicity) of the galactic interstellar medium, the reservoir out of which stars form and supermassive black holes feed.
An observational strategy to study such processes is to focus on the gas at the interface between the star-forming disc and the halo, which contains information on the combination of all the mechanisms described above. 
This gas is often referred to as `extra-planar gas' or, as we prefer to use in this study, `disc-halo interface' (DHI).

The Milky Way is an optimal laboratory to study the DHI, as our privileged position within the Galaxy offers an almost complete all-sky view of this medium.
Observational manifestations of ongoing Galactic gas cycle are the so-called high-velocity and intermediate-velocity clouds \citep[HVCs and IVCs;][]{WakkervanWoerden97}, spatially extended complexes of neutral and ionised gas with line-of-sight velocities incompatible with them participating to the Galactic rotation. 
These features have been extensively studied in \hi\ emission \citep[e.g.,][]{wakker91,wakker07,wakker08,westmeier18} and in absorption of neutral, low- and highly-ionised gas towards bright background QSOs \citep{sembach03,shull09,benbekhti12,lehner12,richter17,clark21} and halo stars \citep{lehner09,lehner11,werk19,Bish+19}.
\hi\ observations have the advantage of providing a uniform coverage across the sky but are sensitive to column densities larger than $\sim10^{18}\cmmq$ \citep[e.g.,][]{lockman02}. 
Absorption measurements instead can probe down to much lower hydrogen columns, but only towards a sparse collection of background targets.
However, unlike absorption-line studies beyond the Local group that are typically limited to about one background QSOs per galaxy \citep[e.g.,][]{tumlinson11,liang14,werk14,werk16}, which makes impossible to characterize individual galaxy halos in any details, in the Milky Way we have the luxury of probing the DHI using several tens of background halo stars with known distance and hundreds of QSOs. 
This makes the Milky Way among the best systems for studying the DHI with absorption-line observations.

Understanding the nature of the HVCs and IVCs requires the knowledge of three key properties: their distance, metallicity and 3D kinematics.
Considerable observational efforts have been devoted to determine the former two properties \citep[e.g.,][]{wakker01,richter01,collins03,tripp03,wakker07,wakker08,thom06,thom08,zech08,lehner10,fox16}. 
The picture that consistently emerges from these studies is that the IVCs are nearby systems ($d\!<\!1$--$2\kpc$) that have near or Solar metallicity, while the HVCs are located at larger distances ($d\!<\!10$--$15\kpc$)\footnote{We do not consider the Magellanic Stream in this classification.} and feature metallicity in the range $0.1$--$1.0\, Z_\odot$.
This dichotomy has been traditional interpreted as an evidence for a distinct origin for the two cloud populations: internal for the IVCs, external for the HVCs.
While the GF is the only plausible internal mechanism that can bring gas at the DHI, several options exist for the external scenario.
These include spontaneous condensation of the hot corona as a consequence of thermal instability, an option that is highly debated amongst the theoretical community \citep[e.g.,][]{Binney+09,Nipoti10, SobacchiSormani19, SormaniSobacchi19}, gas stripped or ejected from satellites \citep[e.g.,][]{Bland-Hawthorn+98,putman06,olano08}, fragments of gas filaments accreting from the cosmic web onto the Galaxy \citep[e.g.,][]{keres09,fernandez12,vandevoort19}, or gas-rich mini-halos \citep[e.g.,][]{GalyardtShelton16}.

Comparatively, the study of the 3D kinematics of the Galactic DHI has not received the same level of attention. 
This is surprising, given that the HVCs and IVCs were originally selected on pure kinematic grounds, and kinematics play a crucial role in the interpretation of their origin and in the measurement of their inflow and outflow rates \citep[e.g.,][]{fox19,clark21}.
\citet{marasco11} were the first to study the global kinematics of the Galactic DHI in the \hi\ phase by applying a parametric model of a thick disc to the data of the LAB Survey \citep{kalberla05}. 
They found evidence for a global rotation with a speed comparable to - but slightly lower than - that of the disc, and for an inflow with a velocity of $30$--$40\kms$.
Similar kinematics are found in the neutral and ionised DHI of external galaxies \citep{Fraternali02,marasco19,li21} and in the Milky Way \siiv-bearing gas \citep{Qu+20}.
The presence of rotation strongly supported the idea of an internal origin for most of the DHI, but the slower rotation speed and the inflow were not trivial to interpret, plus the model did not reproduce several of the high-velocity \hi\ complexes. 

Building on the pioneering works by \citet{fraternali06,fraternali08}, \citet{MFB12} applied a dynamical model of the GF to the same \hi\ dataset.
In this model, fountain clouds are ejected from the disc by supernova feedback and travel through the halo region before returning back to the Galaxy.
Fountain clouds are assumed to be initially ionised but recombination occurs during their orbit, thus their neutral phase is observed preferentially in inflow.
Also, clouds interact with a pre-existing, slow-rotating, metal-poor hot corona, exchanging momentum with it (which slows their rotation down) and triggering its condensation and subsequent accretion onto the disc \citep{marinacci11,armillotta17}. 
This model reproduces the Galactic \hi\ data remarkably well, explains the origin of the slow rotation and global inflow motion, and predicts an accretion of coronal material onto the disc at a rate similar to the Galactic star formation rate.
According to this model, a typical fountain episode produces IVCs that contain only a small percentage (up to $10$--$15\%$ in mass) of condensed coronal gas.
HVCs instead stem from more energetic episodes localised along the Galaxy spiral arms, and can contain a  higher percentage of condensed material \citep[about $50\%$ for complex C;][]{fraternali15}, which explains the wide metallicity spread centred around sub-Solar values that is typically measured in these systems.
Thus, in this framework, the properties of the DHI are explained by a combination of an internal and an external mechanisms, which is in contrast with scenarios where the origin of the HVCs is purely external.

In \citet{lehner22} (hereafter \citetalias{lehner22} of this series), we have built a sample of intermediate- and high-velocity absorption features from the analysis of ultraviolet (UV) spectra of 55 halo stars with high-quality distance measurements.
In the current study (Paper II) we make use of simple, parametric models for the gas distribution and kinematics of the Galactic DHI that we constrain using the dataset built in \citetalias{lehner22}.
We show that the resulting kinematics place strong constraints on the origin of the observed features.

This study is structured as follows.
In Section \ref{s-data} we briefly summarise the main results of \citetalias{lehner22} and describe the absorption dataset that we use in this study.
In Section \ref{s-model} we describe our kinematic models and their application to the data.
The main results of our modelling approach are presented in Section \ref{s-modres} and are discussed in the broader context of the origin of the Galactic DHI in Section \ref{s-disc}.
Conclusions are drawn in Section \ref{s-sum}.

In this work we assume Galactic constants $R_{\odot}\!=\!8.2\kpc$ and $v_{\odot}\!=\!232.8\kms$ \citep{McMillan17}.

\section{Dataset description and basic kinematics}\label{s-data}
We briefly summarise here the main properties of of the absorption line dataset built in \citetalias{lehner22}, along with the main results of that study.
We have targeted a sample of 55 bright (B-type, PAGB/BHB) halo stars at Galactic latitudes $|b|>15^\circ$ using Cosmic Origins Spectrograph (COS) and  Space Telescope Imaging Spectrograph (STIS) on board of the \emph{Hubble Space Telescope} (HST cycle 17 and 20 programs---PIDs 11592 and 12982, plus additional archival HST data).
HST spectra covers the wavelength range between $1150$ and $1730\,\rAA$ with a spectral resolution varying between $R\!\simeq\!17,000$ (for COS) and $R=45,800$--$114,000$ (for STIS), corresponding to a velocity resolution (FWHM) between $2.6$ and $18\kms$.
Using these spectra, we have searched for absorption line features of intervening gas in the several transitions of atomic and ionic species observable in the wavelength range covered, which includes \oi\ $\lambda$1302, \cii\ $\lambda$1334, \civ\ $\lambda\lambda$1548, 1550, \sii\ $\lambda\lambda$1250, 1253, 1259, \siii\ $\lambda$1190, 1193, 1260, 1304, 1526, \siiii\ $\lambda$1206, \siiv\ $\lambda\lambda$1393, 1402, \alii\ $\lambda$1670, \feii\ $\lambda$1608.
The rule to claim for a detection is to observe a given feature in at least two different transitions with the same velocity offset.
Features are classified as high- or an intermediate- velocity clouds (HVCs or IVCs) depending on their flux-weighted mean velocity in the Local Standard of Rest (LSR): we take $|v_{\rm LSR}|\!>\!90\kms$ for the HVCs and $40\le|v_{\rm LSR}|\le90$ for the IVCs.
The deep, often saturated lines at $|v_{\rm LSR}|\!<\!40\kms$ are assumed to be due to the Galactic interstellar medium and are excluded from our analysis.
Contamination from photospheric stellar lines is a potential issue only for $5$ targets which are carefully analysed using a combination of stellar templates and \hi\ emission-line spectra to determine the origin of the features observed.

A major improvement with respect to previous studies \citep{lehner11,lehner12} is the use of Gaia EDR3 parallaxes \citep{gaiadr3} to determine directly the distance of most of the target stars in the sample without relying on models of stellar photospheres. In our sample, most distance determinations with Gaia EDR3 have an accuracy of about or better than $20\%$. All the remaining stars but one are associated with globular clusters that have an accuracy on the distance of about $12\%$.
Our sample spans a distance range $1.4\!<\!d\!<\!19.4\kpc$ (median $d$ of $6.6\kpc$), corresponding to vertical offsets from the Galaxy midplane $0.7\!<\!|z|\!<\!14.2\kpc$ (median $|z|$ of $4\kpc$).
We remark that the resulting dataset is not representative for the whole Galaxy halo, but it probes an approximately spherical region of several kpc around the Sun, with the exclusion of the Galactic disc.

A key result from \citetalias{lehner22} is that the covering factor of target stars that are further away from the Galaxy midplane is similar to that determined for distant QSOs, indicating that our sample spans a sufficiently large range in $|z|$ to encompass most of the DHI traced by UV absorption lines.
This allows us to determine the scale-height of the absorbing gas from the trend of the covering factor with $|z|$.
By assuming a vertical density profile $\propto \sinh(z/h)/\cosh^2(z/h)$, which provides a good description for the extra-planar \hi\ gas of external galaxies \citep{oosterloo+07,marasco19}, we infer $h\!=\!1.4\pm0.2\kpc$ for the whole dataset and $1.0\pm0.3\kpc$ and $2.8\pm0.3\kpc$ for the IVCs and the HVCs, respectively.
Thus IVCs are confined within about a kpc from the disc, while HVCs are located at larger distances, which is consistent with the picture shown in \hi\ HVCs and IVCs using the distance bracket method \citep[e.g.,][]{ryans97,ryans97a,WakkervanWoerden97,wakker01,wakker07,wakker08,thom08}.
We stress that features at $|v_{\rm LSR}|\!>\!170\kms$ (sometimes referred to as very high-velocity clouds, VHVCs) are not observed in our stellar sample but are routinely detected towards QSOs \citep[e.g.,]{sembach03,lehner12,richter17}. This indicates that the VHVC population is located at $|z|\gtrsim10\kpc$, and is possibly associated with the Magellanic Stream and its Leading Arm \citep{lehner12}, or with the outer regions of the Galactic corona \citep{marasco13}.

\begin{figure*}
\includegraphics[width=0.9\textwidth]{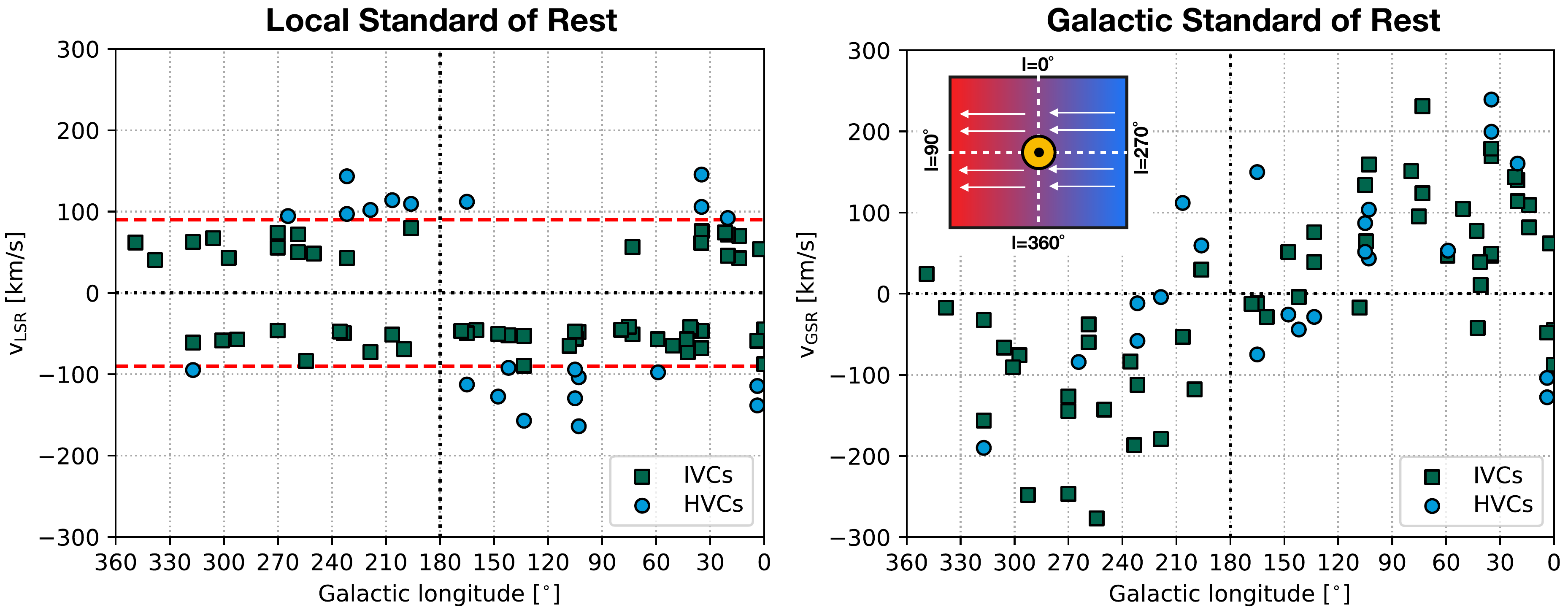}
\caption{Longitude-velocity diagrams for the absorption features studied in this work. IVCs and HVCs are represented by green squares and blue circles, respectively. The \emph{left} panel shows velocities in the LSR, where no significant trends with longitude is visible, suggesting the absence of strong peculiar motions in this reference frame. The \emph{right} panel shows velocities in the GSR. The resulting sinusoidal pattern indicates that the absorbing gas streams along the same direction of the Sun motion, as illustrated in the inset on the top-left.
Both panels convey the same message: the dominant motion of the absorbing material is rotation around the Galactic centre with a speed comparable to that of the disc.}
\label{f-lv}
\end{figure*}

While detailed kinematic models are presented in the next Section, basic considerations on the main kinematic properties of the absorbers in our sample can be done without recurring to modelling.
In Fig.\,\ref{f-lv} we show the longitude-velocity distribution of the features in our dataset.
No significant trends are visible when velocities are determined with respect to LSR (left panel of Fig.\,\ref{f-lv}), indicating that the absorbing gas does not have strong peculiar motions in this particular reference frame.
Alternatively, this property can be visualised by converting velocities to the Galactic Standard of Rest\footnote{Defined as a reference frame where the Sun is at rest with respect to the Galactic centre, $v_{\rm GSR}=v_{\rm LSR}+v_{\odot}\sin(l)\cos(b)$} (GSR), shown by the right panel of Fig.\,\ref{f-lv}: the sinusoidal pattern of the $l$-$v_{\rm GSR}$ diagram, which features a maximum at $l\simeq90^{\circ}$ and a minimum at $l\simeq270^{\circ}$, is consistent with the absorbing gas streaming along the same direction of the Sun motion (that is, towards $l=90^\circ$) as illustrated in the top-left inset of the panel.
Both the IVCs and the HVCs seem to participate to this global pattern, although the trend is more defined for the former than for the latter.

These considerations indicate that the absorbing gas rotates in the same direction of the Galaxy, and that such motion is the dominant component of its kinematics \citep[see also][]{Qu+20}.
This contrasts with the assumption of pure vertical motions that is often adopted to determine gas inflow and outflow rates in the Galaxy \citep[e.g.][]{fox19,clark21}.
We stress that the change of reference frame from LSR to GSR helps to provide a clearer visualisation of this rotating pattern but does not add further information on the 3D kinematics of the gas, which is the subject of the analysis below.

\section{Kinematic models}\label{s-model}
Our models stem from those implemented by \citet[][hereafter \citetalias{marasco19}]{marasco19} to describe the morphology and kinematics of the \hi\ in the halo of nearby galaxies from the HALOGAS Survey \citep{heald11}.
Similar models have been used by \citet{marasco11} to describe the properties of the \hi\ in the halo of the Milky Way.
We briefly summarise here their features, along with the adjustments required to adapt them to the current study.

\subsection{Model details}\label{s-model-details}
In \citetalias{marasco19}, the gas layer was modelled as an axi-symmetric, smooth gas distribution described by four kinematic parameters, namely the gas velocity vector in cylindrical coordinates ($v_{\rm R}$, $v_\phi$, $v_{\rm z}$) plus the gas velocity dispersion $\sigma$, and three structural parameters defining the gas radial and vertical distribution. 
However, due to the intrinsic difficulties in modelling a sparse collection of absorbing features, in this study we focus solely on the large-scale gas kinematics and fix the gas structure and velocity dispersion as follows.

First, we assume a surface density profile that is constant with radius.
This choice is driven by the fact that, since most background sources are high-latitude halo stars with a wide $|z|$ distribution, our data are very informative on the gas profile in the vertical direction but contain little information along the radial direction. We tested this by attempting to fit the data (see Section \ref{s-modelfit}) with a more complex model featuring the same the parametric surface density profile adopted by \citetalias{marasco19} (see their eq.\,(1)), finding that we could not constrain the parameters describing the surface density.
Secondly, we use the scale-height $h$ determined for the DHI in \citetalias{lehner22} on the basis of the trend of the covering factor with $|z|$, as discussed in Section \ref{s-data}.
This is a necessary assumption given the smooth nature of our DHI model.
Finally, we set $\sigma$ to the value of $20\kms$. 
This value is slightly larger than the typical line-widths of our absorbing features ($10\!-\!15\kms$) but, by visually comparing the model with the data (see Section \ref{s-modres} and Figs.\,\ref{f-bestmodel-XVCs} and \ref{f-bestmodel-inout}), we found it to provide a better description for the velocity spread of our dataset. 
We have verified that assuming a different value of $\sigma$ has a negligible impact on our results.
This leaves us with three free parameters, which are those describing the velocity vector ($v_{\rm R}$, $v_\phi$ and $v_{\rm z}$).

For any given choice of these three parameters, a Monte Carlo sampling is used to generate the distribution of the ionised gas in the full 6D phase-space. An `internal' projection of such distribution is derived by considering a reference frame located at ($R\!=\!R_\odot$, $z\!=\!0$), rotating in the azimuthal direction with a speed equal to $v_\odot$. This frame is used to transform the 6D phase-space into a 4D `observational' space ($l$, $b$, $v_{\rm LSR}$, $d$), which can be compared with the absorption data as we discuss below. To simplify our notation, in what follows we refer to the velocity measured in the LSR as $v$, rather than $v_{\rm LSR}$.

\subsection{Fitting the model to the data}\label{s-modelfit}
Comparing the smooth, all-sky gas distribution resulting from our model with the discrete absorption line dataset built in this study is not a trivial task. Ideally, for any given sight-line probed by a background star, one would derive model absorption line profiles of the various ionised species observed for comparison with the data. This would demand a detailed knowledge of both the physical conditions of the gas and the ionising radiation field, which clearly are not readily available. In this study, instead, we adopt a simplified, purely kinematic approach based on the philosophy that the optimal model is the one that best reproduces the distribution of the absorbers in the observed ($l$, $b$, $v$, $d$) space. To pursue this approach, we first need to build the observed and predicted distributions, and then to define the method to compare them.

\begin{figure*}
\includegraphics[width=0.9\textwidth]{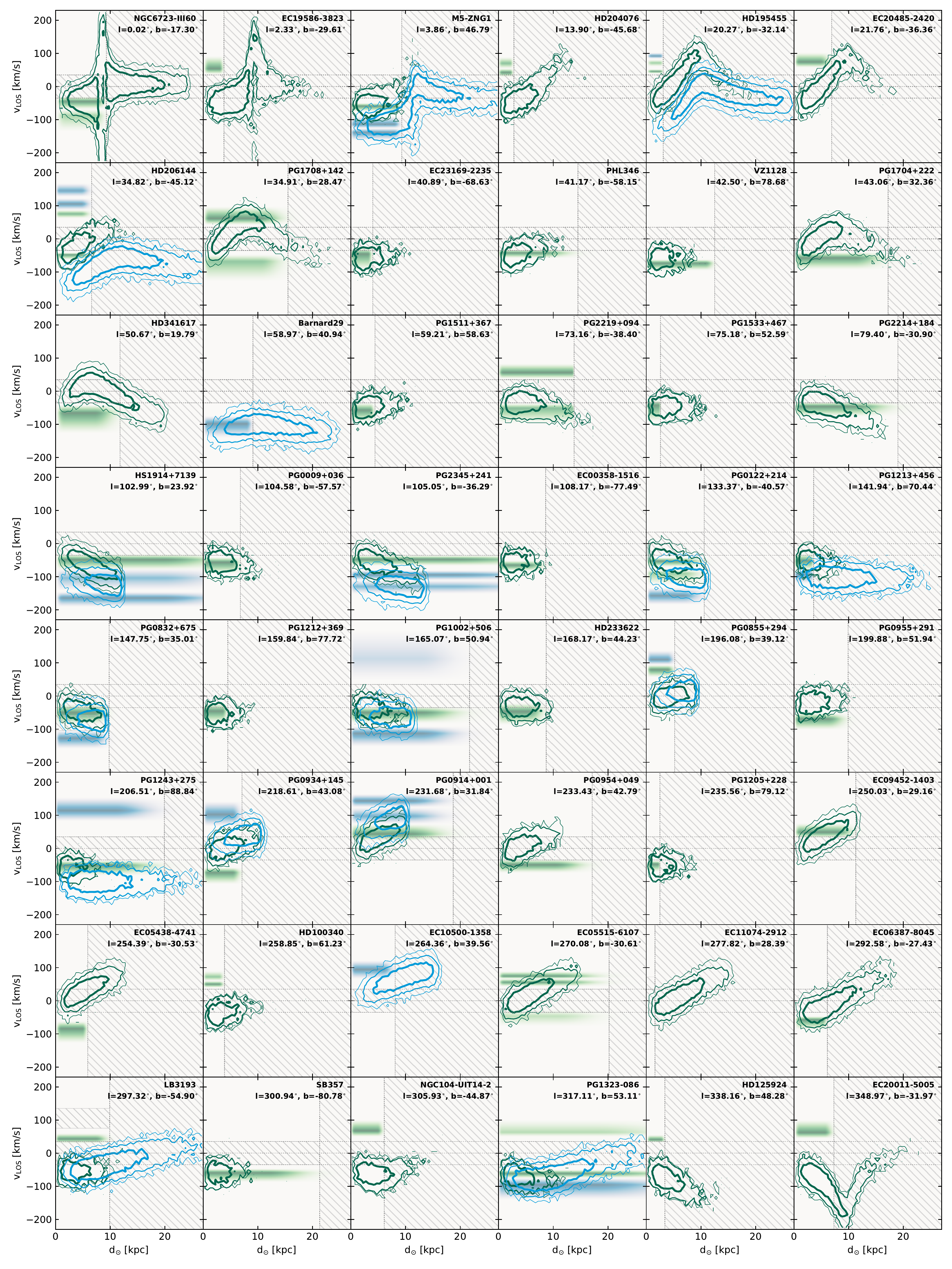}
\caption{Comparison between the kinematics of the ionised absorbers around the Milky Way and those predicted by our two-component (IVCs + HVCs) model. Each panel shows the ($v$, $d$) probability distribution towards a given target star, indicated in the top-right corner. The data are shown as shaded horizonthal stripes (in green for the IVCs, in blue for the HVCs), with thicknesses corresponding to the velocity widths of the absorption features. Our best-fit models are shown with green (IVCs) and blue (HVCs) contours, representing the $68\%$, $95\%$ and $99.7\%$ of the enclosed probability. The hatched areas highlight regions at distances or velocities that cannot be probed by the observations.}
\label{f-bestmodel-XVCs}
\end{figure*}

\begin{figure*}
\includegraphics[width=0.9\textwidth]{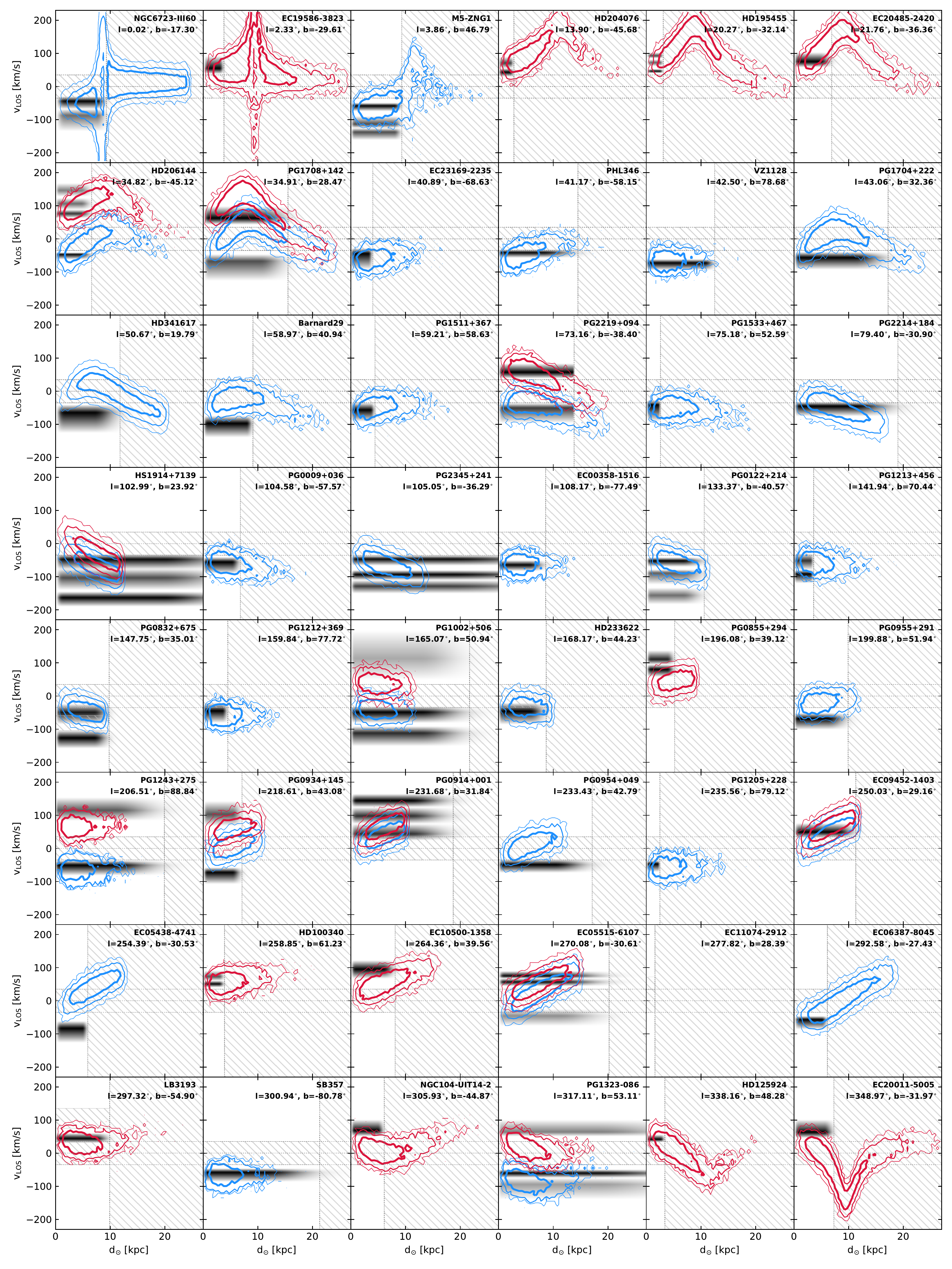}
\caption{Comparison between the kinematics of the ionised absorbers around the Milky Way and those predicted by our two-component (inflow+outflow) model. Each panel shows the ($v$, $d$) probability distribution towards a given target star, indicated in the top-right corner. The data are shown as shaded horizonthal stripes, with thicknesses corresponding to the velocity widths of the absorption features. Our best-fit models are shown with blue (inflow) and red (outflow) contours, representing the $68\%$, $95\%$ and $99.7\%$ of the enclosed probability. The hatched areas highlight regions at distances or velocities that cannot be probed by the observations.}
\label{f-bestmodel-inout}
\end{figure*}

We focus first on our model. For each target halo star (Table 1 in \citetalias{lehner22}), we produce a synthetic lightcone by considering all Monte Carlo sampler (or `clouds') within a fixed aperture $\delta$ from the ($l$,$b$) sight-line of the star. These clouds are used to build a smooth probability distribution in the ($v$, $d$) space relative to that particular sight-line. In principle, $\delta$ should sample an angle similar to the background stars' angular size (micro-arcsecond scales), 
%
%
but values up to $\sim8^{\circ}$ have a negligible impact on our results and provide a better sampling of the probability distribution for a fixed number of clouds $N_{\rm cl}$. After experimenting with different values of $N_{\rm cl}$ and $\delta$, we found that $N_{\rm cl}\!=\!2.5\times10^6$ and $\delta=4^{\circ}$ give the optimal compromise between model accuracy and computational speed.
We stress that these $N_{\rm cl}$ samplers are by no means representative of realistic clouds in the Galactic halo, but are simply used as a convenient way to sample a continuous gas distribution.
Examples of model probability distribution associated with the various sight-lines are shown in Figs.\,\ref{f-bestmodel-XVCs} and \ref{f-bestmodel-inout} as solid lines. Such distributions will depend on the model parameters: for instance, larger scale-heights will stretch them towards higher $d$, while negative $v_{\rm z}$ will shift them towards more negative $v$ (especially for high latitude sight-lines), and so on.

We must now build the corresponding distribution for the data. Unfortunately, the data do not fully sample the ($v$, $d$) space but provide only discrete sampling in the velocity direction, corresponding to the $v_{\rm LSR}$ entries reported in Table 2 of Paper I. We therefore approximate the distribution in the $v$ direction by considering a triangular function that starts from (and ends at) zero at velocities equal to $v_1$ and $v_2$ (as reported in Table 2 of paper I), and reaches a maximum at the observed $v_{\rm LSR}$. We assume a constant probability distribution in the $d$ direction for $d<d_\star$ ($d_\star$ being the distance of the background star), with a decline at larger $d$ following a Gaussian distribution with standard deviation given by the uncertainty on $d_\star$. We normalise the distribution associated with each feature by setting to one the probability integrated in the ($v$, $d$) space. 
In Figures \ref{f-bestmodel-XVCs} and \ref{f-bestmodel-inout}, the observed probability distributions are shown as shaded horizontal stripes.

A `likelihood' estimator $\mathcal{L}$ can be determined by multiplying the probability distribution of the model by that of the data. With our implementation, this can be conveniently computed as
\begin{equation}\label{e-like}
    \mathcal{L}\propto\frac{1}{n_{\rm tot}} \sum_{i=1}^{N_\star} \left(\sum_{n=1}^{n_{\rm cl}} \mathcal{P}_i(v_n, d_n)\right)
\end{equation}
where $n_{\rm tot}$ is the total number of clouds, the external sum is extended to all $N_\star$ sight-lines, the internal sum is extended to all $n_{\rm cl}$ clouds along a given sight-line $i$, $\mathcal{P}_i(v_n, d_n)$ is the observed probability density for sight-line $i$ computed at the ($v_n$,$d_n$) location of cloud $n$. In practice, eq.\,(\ref{e-like}) defines $\mathcal{L}$ as proportional to the sum of the observed probability density sampled at all the ($v$, $d$) locations of the model clouds, which mathematically corresponds to multiplying the observed and the predicted probability distributions, as requested. The value of $\mathcal{L}$ will increase (decrease) when more (less) clouds fall within ($v$, $d$) regions compatible with our data, while the $n_{\rm tot}$ normalisation factor in eq.\,(\ref{e-like}) ensures that $\mathcal{L}$ does not depend on the total number of clouds used to build our model.

In our computation of $n_{\rm tot}$ we do not account for clouds located in `forbidden' regions of the ($v$, $d$) space that cannot be probed by the observations. Specifically, we exclude clouds with $|v|\!<\!40\kms$, as this velocity regime is systematically occupied by material in the Galaxy disc\footnote{The velocity range occupied by the disc varies with $l$ and $b$, but the adopted constant velocity cut is a very good approximation for $|b|\!>\!30^{\circ}$.}, along with clouds at distances beyond those of the target stars (plus the distance error) and those falling in ($v$, $d$) regions associated with features contaminated by stars ($f_\star=1$ in Table 2 of \citetalias{lehner22}).
Sightlines with non-detections are included in the analysis since, even though they do not contribute to increase the probability sums in eq.\,(\ref{e-like}) (since $\mathcal{P}_i\!=\!0$ everywhere), they  affect the value of $n_{\rm tot}$ if clouds are present along those directions.

To determine the optimal model parameters, the likelihood is multiplied by an uninformative (flat) prior to give the posterior probability distribution, which is sampled with an affine-invariant Markov Chain Monte Carlo (MCMC) method using the \texttt{python} implementation by \citet{emcee}. The MCMC method allows us to fully sample the parameter space in order to account for possible degeneracies, which may be expected given the limitations of our dataset. We stress that, while the expression defined in eq.\,(\ref{e-like}) is a reasonable `figure of merit' function to compare the model with the data, it is extremely hard to robustly quantify the uncertainties associated with our modelling. This is primarily due to the intrinsic differences between the model, which represents a smooth and axi-symmetric distribution of gas, and the observations, which are a sparse collection of absorption line measurements. We therefore expect our MCMC routine to capture the overall shape of the posterior distribution, but to be agnostic on its width, which implies that the estimated uncertainties on the model parameters must be considered with caution.
 
\section{Results of the models} \label{s-modres}
\begin{table}
\caption{Best-fit parameters for our two-component models of ionised gas in the Galactic halo. The model likelihoods are reported in the last row.}
\label{t-modparam} 
\centering
\begin{tabular}{ccccc}
\hline\hline
\multicolumn{1}{c}{parameter} & \multicolumn{2}{c}{Scenario 1} & \multicolumn{2}{c}{Scenario 2}\\
\hline
    & IVCs & HVCs & Inflow & Outflow\\
    & (\km) & (\km) & (\km) & (\km)\\
\hline
$v_\phi$    & $224\pm19$  & $143^{+67}_{-33}$ & $233\pm18$  & $232^{+91}_{-54}$\\
$v_{\rm z}$ & $-57^{+18}_{-13}$  & $-102^{+33}_{-59}$ & $-67^{+8}_{-11}$  & $62^{+55}_{-40}$ \\
$v_{\rm R}$ & $25^{+15}_{-20}$   & $63^{+45}_{-53}$ & $30^{+15}_{-11}$   & $-22^{+39}_{-35}$\\
\hline
\multicolumn{1}{c}{$\log{\mathcal{L}}$} & \multicolumn{2}{c}{$-203.16$} & \multicolumn{2}{c}{$-287.65$}\\
\hline
\end{tabular}
\end{table}

\begin{figure*}
\includegraphics[width=\textwidth]{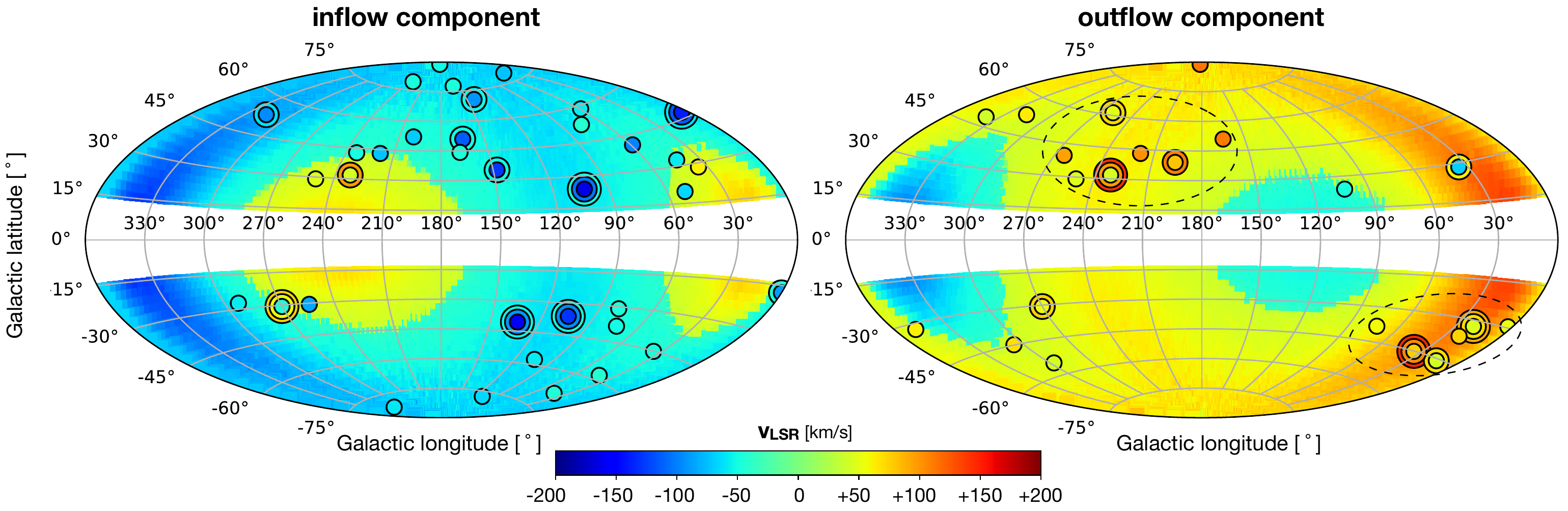}
\caption{Comparison between the all-sky velocity fields predicted by our best-fit inflow+outflow models (background colours) and the LSR velocity of the observed absorption features (coloured circles). Concentric circles are used when multiple detections are found towards the same target star. The map on the left (right) shows the 52 (33) absorbers compatible with the inflow (outflow) component. The dashed ellipses mark the locations of the bi-conic outflow encompassing most of the observed feature associated with the outflow component.}
\label{f-modelmaps}
\end{figure*}

We run our MCMC routine using 64 walkers and chains of 2000 steps.
After visual inspection of the chains, we decided to discard the initial $500$ `burn-in' steps beyond which the chains stabilise, and proceeded to inspect the posterior probability distributions.
We found the posterior to be multi-modal, featuring three separate peaks in the ($v_{\rm R}$, $v_\phi$, $v_{\rm z}$) parameter space. 
This indicates that a single-component model provides a poor representation of our data, and that a more complex, multi-component model is required to fully account for the data complexity.
We investigate this further below, exploring two different scenarios.

\subsection{Scenario 1 - IVCs and HVCs}\label{ss-scenario1}
In this scenario, we consider the IVCs ($|v|\le90\kms$) and the HVCs ($|v|\!>\!90\kms$) as separate populations, each described by its own scale-height \citepalias[fixed to $1.0\kpc$ for the IVCs and to $2.8\kpc$ for the HVCs, see Section \ref{s-data} and][]{lehner22} and kinematics.
We stress that this is a somewhat phenomenological separation, based on the classical distinction between the HVCs and IVCs, and makes no assumption on the physical origin of these two populations.

We fit each population separately, finding a well defined posterior distributions for the IVCs and a double-peaked posterior for the HVCs. These are shown in the top panels of Fig.\,\ref{f-corner}.
We take the median of such distributions as best-fit values, and compute the error-bars as half the difference between the 84th and 16th percentiles. 
The parameters derived, reported in the third and fourth columns of Table \ref{t-modparam}, indicate that the IVC population is characterised by fast rotation velocities (compatible with $v_\odot$) and an inflow speed of $\sim60\kms$, whereas the HVC population rotates slower ($120-200\kms$) and accretes at a faster pace ($\sim100\kms$).
We get positive $v_{\rm R}$ for both components, but with uncertainties that are compatible with $0$.
Interestingly, the different inflow velocities are compatible with a scenario where the HVCs `rain down' from the CGM and get decelerated in the process by interaction with the hot corona, finally appearing as IVCs when they approach the Galactic disc.
This would also explain the different scale-heights found for the two populations.

Our best-fit model and our dataset are compared to each other in Fig.\,\ref{f-bestmodel-XVCs} that shows, for all the sight-lines that feature at least one detection, the probability distribution of the model (green and blue contours for the IVC and HVC component, respectively) and of the data (green-shaded and blue-shaded horizontal stripes for the IVCs and HVCs, respectively) in the ($v$, $d$) space.
The IVC (HVC) model is shown only if IVC (HVC) features are detected along a given sight-line.
Clearly, our model seems to be compatible with the $v$-$d$ distribution of the absorption features observed towards several target stars. However, it fails to reproduce many of the features observed at positive $v$ (see, for instance, the third-to-sixth panels in the top row of Fig.\,\ref{f-bestmodel-XVCs}, or the first panel in the second row). 
In total, out of $78$ features, $20$ ($6$ HVCs, $14$ IVCs) are clearly incompatible with our model, indicating that a scenario where the IVCs and HVCs are treated as separate components, each described by its own simple kinematics, is not fully adequate to describe the observed dataset.

\subsection{Scenario 2 - Inflow and Outflow}\label{ss-scenario2}
In the second scenario, we adopt a more physically motivated distinction and model the DHI as a combination of an outflow component ($v_{\rm z}\!>\!0$), representing material ejected from the Galactic disc by stellar feedback, and an inflow component ($v_{\rm z}\!<\!0$) either due to the previously ejected material returning back to the disc, or to new gas that joins the Galaxy for the first time.
This approach adds further complexity, as we do not know a priori which absorption feature is associated with which component.
Our strategy is to fit both components at the same time, assigning each sight-line to one component or the other depending on which provides the highest `local' likelihood, given by the term in parenthesis in eq.\,(\ref{e-like}).
We further consider the possibility that both components are visible along a given sight-line, which we adopt when the local likelihood of the combined inflow+outflow model is larger than $f\times$ the highest single component likelihood.
We set $f$ to $1.2$ (that is, we require a $20\%$ better likelihood to accept both components at the same time), which we found to optimally capture the few occurrences of this situation.
With this approach, the assignment of a given sight-line to either component (or to both components simultaneously) will vary depending on the model parameters, which are fit to the data.
For simplicity, we fix the scale-height of both components to the value of $1.4\kpc$, representative for the whole ionised gas layer as discussed in Section \ref{s-data}.

The best-fit parameters obtained for this scenario are listed in the rightmost two columns of Table \ref{t-modparam}.
Here, the ionised gas is best described by two rapidly rotating ($v_{\phi}\sim v_\odot$) gas layers, one inflowing and the other outflowing at about the same speed ($\sim65\kms$).
As in the previous scenario, also in this case radial motions are present, but are small ($\sim30\kms$) for the inflow component and compatible with zero for the outflow component.
With respect to the inflow, the parameters of the outflow are characterised by large uncertainties, which are mainly caused by the posteriors of $v_\phi$, $v_{\rm z}$ and $v_{\rm R}$ being positively correlated, as shown in the bottom panels of Fig.\,\ref{f-corner}.
Compared to the previous case (Fig.~\ref{f-bestmodel-XVCs}), this scenario provides a much better likelihood, reported in the last row of Table \ref{t-modparam}.

This model and the data are visually compared in Fig.\,\ref{f-bestmodel-inout} using the ($v$, $d$) diagrams, as in the previous case.
Clearly, this scenario is better suited to reproduce the kinematics of the absorbers, as virtually all features show some degree of overlap with at least one of the two components in the ($v$, $d$) space.
In particular, $45$ features ($58\%$) are compatible with the inflow component alone, $26$ ($33\%$) with the outflow, and $7$ ($9\%$) with both at the same time.
The highest number of features reproduced by the inflow can be interpreted as different covering fraction associated the two components: that is, the inflowing gas is more diffuse whereas the outflowing material is more clumpy and/or collimated.
This can be better appreciated in Fig.\,\ref{f-modelmaps}, which shows the all-sky moment-1 map\footnote{derived in our model by excluding gas at $|v|<40\kms$ and $d\!>\!7\kpc$, corresponding to the median $d$ of our target star sample} predicted by our inflow (on the left) and outflow (on the right) model, overlaid by the absorption features associated with each component.
We stress that this visualization is less informative than the ($v$, $d$) diagrams of Fig.\,\ref{f-bestmodel-inout}, since it misses the information along the line of sight.
However, it helps in revealing that the majority ($70\%$) of the features associated with the outflow occupy two well-defined regions in the sky (dashed ellipses in Fig.\ref{f-modelmaps}).
Such regions have approximately the same $|b|$ (but with opposite sign) and are separated by $\Delta l\simeq180^\circ$, suggesting the presence of a bi-conic outflow driven by star formation in the vicinity of the Sun.
This is a very intriguing feature. 
Biconic outflows of ionised gas are routinely observed at the centre of starbursts or in galaxies hosting AGN, but with speeds that can reach thousands $\kms$, \citep[e.g.,][]{Veilleux05,fiore+17}.
Outflow velocities of about $50$--$100\kms$, comparable with our estimate for $v_{\rm z}$, can instead be due to an expanding superbubble or to a particularly prominent star forming region \citep[e.g.][]{kim18}.
Unfortunately, pinpointing to the exact location in the disc where the ejection is supposed to occur is not straightforward and would require a dedicated dynamical investigation.

\section{Discussion}\label{s-disc}
\subsection{Relation to the galactic fountain mechanism}\label{ss-gf}
The results of the previous Section indicate that the observed absorption features are well described by two gaseous components, both rotating with a speed similar to that of the Galactic disc, accreting onto or escaping from the Galaxy with vertical speeds of several tens $\kms$.
The dominant kinematics of either component is the rotation at $v_\phi\sim v_\odot$, indicating that both components are ``aware'' of the presence of the disc and, possibly, are causally related to it.
Radial motions, if present, are limited to a few tens $\kms$ (for the inflow) or are affected by large uncertainty (for the outflow).
A promising physical mechanism that can explain such kinematics is that of the GF \citep[e.g.,][]{shapiro76,bregman80,fraternali06}, where gas is expelled from star-forming regions by stellar feedback and travels through the lower regions of the halo before making its return to the disc. 

The main argument against a GF origin is given by the excess of features associated with the negative $v_{\rm z}$ component, which is not trivial to interpret.
Dynamical models of the Galactic \hi\ at the DHI indicate that the ouflowing leg of the fountain must be largely ionised \citep{MFB12}, thus one would expect the majority of the ionised absorbers to be compatible with a model with positive $v_{\rm z}$. Instead, out of $78$ detections, only a number between $26$ and $33$ ($33\%$--$42\%$) are compatible with an outflow.
A similar evidence for prevalent inflow has been found by \citet{Bish+19} by studying the ionised gas absorption features in the spectra of $54$ high-latitude ($|b|\!>\!60^\circ$) blue horizontal branch stars in the Galaxy halo, and by \citet{clark21}, although this study is affected by the caveat discussed in Section \ref{ss-accr}. 
Also, the Milky Way is not the only galaxy where the ionised DHI is seen preferentially in accretion. 
The extra-planar H$\alpha$-emitting gas of the nearby galaxy NGC\,2403 appears to be globally inflowing onto the disc, like its \hi\ counter-part \citep{Fraternali+04}.
\citet{Zheng+17} studied the DHI of M33 in absorption towards $7$ UV-bright stars evenly distributed across the galaxy disc, finding evidence for a widespread inflow.
\citet{li21} modelled the H$\alpha$ emission-line cube of NGC\,3982 and NGC\,4152, two nearby star-forming galaxies, finding that the ionised DHI of these systems, as a whole, is better reproduced by an inflow, rather than by an outflow.

However, our findings of Section \ref{s-modres} provide a very solid counter-argument, based on the variation of the filling factor: the more diffuse (large filling factor) inflow component is easier to detect than the more collimated (small filling factor) outflowing gas.
This would explain not only why inflow features are more routinely observed, but also the results of \citet{li21} based on spatially-resolved H$\alpha$ emission-line observations, as an inflow model will be typically preferred when the data are described with a single component.
Our argument is further supported by the recent hydrodynamical model of stellar feedback developed by \citet{kim18}, who found that the warm gas is mostly confined in small cloudlets during the outflow-dominated phase of the fountain, but becomes the main volume-filling component during the inflow-dominated phase (see their Fig.\,$2$).

A possible complication to this picture is given by the fact that outflows are more routinely observed than inflows in down-the-barrel studies of star-forming galaxies at intermediate redshift \citep[e.g.][]{rubin14}. 
However, down-the-barrel measurements are mainly sensitive to gas in front of UV-bright star-forming regions of the disc, where outflows are presumably more frequent, and may miss any intervening component that is more sparsely distributed over the whole galaxy area. 
This limitation is not present in our work or in emission-line studies like that of \citet{li21}, which are not biased towards regions with high star formation rates.

In the light of these considerations, we argue that our data are largely consistent with a GF scenario where the fountain material is ejected from the disc in collimated structures and returns back as a more diffuse, volume-filling `rain'.

\subsection{Do HVCs and IVCs trace gas accretion onto the Galaxy?}\label{ss-accr}
Star-forming galaxies have required throughout cosmic time a continuous supply of gas to replenish the material used for their star formation \citep[e.g.,][]{PezzulliFraternali16,SchonrichMcMillan17}.
In the local Universe, accretion provided by wet mergers account at most by one-fifth of the galaxy star formation rates \citep{sancisi+08,diteodorofraternali14}, thus most gas accretion must come directly from cold filaments or from the cooling of the hot CGM \citep[e.g.,][]{keres05, dekel06}.
A key question to address is whether some of the ionised features studied in this work are associated with a population of accreting extragalactic clouds, which can potentially refuel the Galaxy at the rate required.
To fully address this question, one would need to know not only the 3D kinematics of the ionised material, but also the hydrogen masses (or column densities) of each absorption feature. This would require detailed photoionisation modelling which goes beyond the scope of this work.
Nonetheless, qualitative considerations can be done.

Recently, \citet{fox19} used the UV metal-line absorption dataset of QSOs built from the COS archive by \citet{richter17} to determine the rates of gas flow around the Galaxy.
They found evidence for both inflow and outflow, as expected from an ongoing GF cycle, and measured an inflow rate larger than the outflow rate, which they took as an evidence for gas accretion onto the Galaxy.
\citet{clark21} used stacked spectra to infer the spatial distribution of the inflow and outflow features, finding that the former were confined to small, well-defined structures, while the latter were spread more uniformly across the sky, which is the opposite of what we find here.
There are three main differences between the present study and the works of \citet{fox19} and \citet{clark21}.
The first is that they focus on the $|v_{\rm LSR}|>90\kms$ velocity range, thus IVCs are excluded from their analysis.
Second, they use QSOs instead of stars, which probe not only the HVCs, but also the VHVC population, which is not observed towards halo stars and is likely to have a distinct origin (\citetalias{lehner22}).
The third difference is that, in their works, inflow and outflow features are separated on the basis of the sign of $v_{\rm GSR}$, rather than using kinematic models as we do in our study.
We caution that this approach may easily lead to spurious results, especially in a scenario where the main motion of the absorbing gas is rotation around the Galactic disc, as we claim here. 
The right-hand panel of Fig.\,\ref{f-lv} exemplifies this problem: adopting the $v_{\rm GSR}$-based separation on our data would result in having half of the sky - the half that the Sun is approaching due to its rotation around the Milky Way centre - populated by outflowing gas, and the other half populated by inflowing gas, which would appear quite unrealistic.
Only for features at very high Galactic latitudes is the line-of-sight velocity (in either the LSR or the GSR frame) a direct measurement of the velocity component perpendicular to the Galaxy disc.

In general, we cannot rule out that some of the features in our dataset are associated with genuine extra-galactic gas accretion, but the results of Section \ref{s-modres} indicate that the bulk of the IVC and HVC population do not originate from a continuous rain of material from outside the Galaxy.
Instead, IVCs and HVCs appear to be manifestations, at different $v_{\rm LSR}$, of the gas cycle occurring at the disc-corona interface triggered by stellar feedback.
That the HVCs too participate to this fountain cycle agrees with the results by \citet{fraternali15} and \citet{marascofraternali17}, who successfully explained the properties of two well studied high-velocity \hi\ complexes -- complex C \citep[e.g.,][]{wakker91} and the Smith Cloud \citep{smith63} -- using a model of the GF featuring high-speed ($\sim200\kms$) outflows from the Galaxy spiral arms.
Interestingly, the large uncertainties associated with the parameters of outflow component in Table \ref{t-modparam} can be interpreted as due to the existence of a wide distribution of outflow velocities, which are expected by theoretical models of stellar feedback, rather than to a unique value of $v_{\rm z}$.

A revision of the standard fountain model, described in detail by \citet{fraternali17}, predicts that the interaction between the fountain material launched from the disc and the coronal gas pre-existing in the halo triggers the condensation of the latter, stimulating its accretion onto the Galaxy.
Because of the momentum exchange between the fountain clouds and the corona, this mechanism is expected to leave subtle signatures in the kinematics of the fountain material, which slows down their rotation and acquires a radially inward motion.
At a fixed coronal gas kinematics, the magnitude of these signatures depends solely on the efficiency of the coronal gas condensation, which therefore can be inferred by applying a dynamical model of the GF to high-quality DHI data.
\citet{MFB12} applied this model to the all-sky \hi\ data of the Milky Way and found that, for a coronal gas accretion rate similar to the Galactic star formation rate, it provided an excellent description of the \hi\ dataset.
\citet{marasco13} showed that the same model reproduced also the global kinematic trend of the ionised gas absorbers observed towards QSOs and towards the small sample of stellar targets available at that time.

While the ascending and descending $v_{\rm z}$ inferred for our dataset are similar to those derived for the Milky Way and for external galaxies by GF models, the kinematic signatures expected in the case of interaction with the corona (decreased $v_\phi$ and negative $v_{\rm R}$ for the descending phase) do not seem to be evident.
Taken at face value, the positive $v_{\rm R}$ found for the inflow component and the $v_{\phi}$ that closely matches that of the disc suggest that the cloud-corona interaction is very limited or does not occur at all.
Unfortunately, as discussed in Section \ref{s-modres}, it is difficult to determine the error-bars associated with our model parameters.
Lowering the value of $v_\phi$ by $10$--$20\kms$ and that of $v_{\rm R}$ by $\sim30$--$40\kms$ gives a model that is slightly worse than that shown in Fig.\,\ref{f-bestmodel-inout}, but still provides a good description of the absorption feature kinematics.
Our conclusion is that, with only the dataset analysed in this study, it is not possible to robustly assess the magnitude of the interaction between the fountain clouds and the CGM.

Finally, we remark that cold-mode accretion occurring at the periphery of the disc \citep[e.g.,][]{Trapp+21} may be another viable gas accretion channel, though virtually impossible to test with the available dataset given the lack of low-latitude target stars.

\section{Conclusions}\label{s-sum}
The disc-halo interface (DHI) of star-forming galaxies is thought to be composed by the combination of processed gas, expelled from the interstellar medium in response to feedback processes from star formation and active galactic nuclei, and fresh material that accretes from the cosmic web.
The characterisation of the properties of this gas layer is a necessary step in the study of the gas cycle around galaxies, which in turn is key to understand galaxy evolution.

In \citetalias{lehner22}, we have built a large dataset of high- ($|v|\!>\!90\kms$) and intermediate- ($40\!<\!|v|\!<\!90\kms$) velocity UV absorption features from the analysis of \emph{HST} spectra of 55 Galactic halo stars with known distance measurements from the Gaia DR3 parallaxes.
These distances were used to study the variation of the covering fraction with $|z|$, from which the vertical distribution of the absorbing gas could be derived.
In this study (Paper II), we have modelled the kinematics of the absorbing gas by comparing the distributions of the observed features in the phase-space (that is, the $v$-distance space) with those predicted by a simple model of a thick disc with $3$ free kinematical parameters, namely the gas rotational, vertical and radial velocities ($v_\phi$, $v_{\rm z}$, $v_{\rm R}$).
Our results can be summarised as follows.
\begin{enumerate}
\item At least two separate components are required to reproduce the data. A single-component model features multiple peaks in the parameter posteriors, suggesting the need of multiple components.
\item A double-component model where the HVCs and the IVCs are treated as separate populations, each with its own parameter set, provides only a partial description of the data as it fails to reproduce several features at $v_{\rm LSR}>0$ (Fig.\,\ref{f-bestmodel-XVCs}).
\item The data are best described by a double-component model made by a combination of an inflow and an outflow, both characterised by rotation with $v_{\phi}\!\simeq\!v_{\odot}$ and vertical speeds of $50\!-\!100\kms$ (Fig.\,\ref{f-bestmodel-inout}). 
\item Most of the features associated with the outflow component are spatially confined towards two opposite direction in the sky, ($l\!=\!220^{\circ}$, $b\!=\!+40^{\circ}$) and ($l\!=\!40^{\circ}$, $b\!=\!-40^{\circ}$), indicating the presence of a bi-conic outflow driven by stellar feedback in the Solar vicinity. The features associated with the inflow are instead spread more uniformly across the sky (Fig.\,\ref{f-modelmaps}).
\end{enumerate}
Our results suggest that the classical, velocity-based separation between HVCs and IVCs is phenomenological as it is not associated with physically distinct kinematics for the two cloud populations.
Instead, they support a scenario where both the HVCs and the IVCs are manifestations, at different velocities, of the same cycle of gas triggered by stellar feedback: that is, the galactic fountain.
While our results refer specifically to the tenuous gas traced by UV absorption lines, which is the subject of this study, similar conclusions can be drawn from the dynamics of the extra-planar \hi\ of our Galaxy \citep{MFB12,fraternali15,marascofraternali17}.
These considerations further highlight the importance of kinematic and dynamical models as necessary tools to infer the properties of gaseous halos.

A corollary of our findings -- also applicable to studies of the DHI in external galaxies -- is that if the ascending and descending phases of the galactic fountain are systematically characterised by different filling factors, the latter will be observed more frequently and consequently be considered as an evidence for gas accretion onto the galaxy, possibly leading to an overestimation of the actual accretion rate.

\section*{Acknowledgements}
AM acknowledges the support by INAF/Frontiera through the "Progetti Premiali" funding scheme of the Italian Ministry of Education, University, and Research. Support for this research was initially and partially provided by NASA through grant HST-GO-12982 from the Space Telescope Science Institute, which is operated by the Association of Universities for Research in Astronomy, Incorporated, under NASA contract NAS5-26555. Based on observations made with the NASA/ESA Hubble Space Telescope, obtained from the data archive at the Space Telescope Science Institute. STScI is operated by the Association of Universities for Research in Astronomy, Inc. under NASA contract NAS 5-26555. This work has made use of data from the European Space Agency (ESA) mission {\it Gaia} (\url{https://www.cosmos.esa.int/gaia}), processed by the {\it Gaia} Data Processing and Analysis Consortium (DPAC, \url{https://www.cosmos.esa.int/web/gaia/dpac/consortium}). Funding for the DPAC has been provided by national institutions, in particular the institutions participating in the {\it Gaia} Multilateral Agreement.

\section*{Data Availability}
The data underlying this article are available in \citetalias{lehner22} and in its online supplementary material.


\bibliographystyle{mnras}
\bibliography{ms} 

\appendix
\section{Corner-plots for the models}
\begin{figure*}
\includegraphics[width=0.84\textwidth]{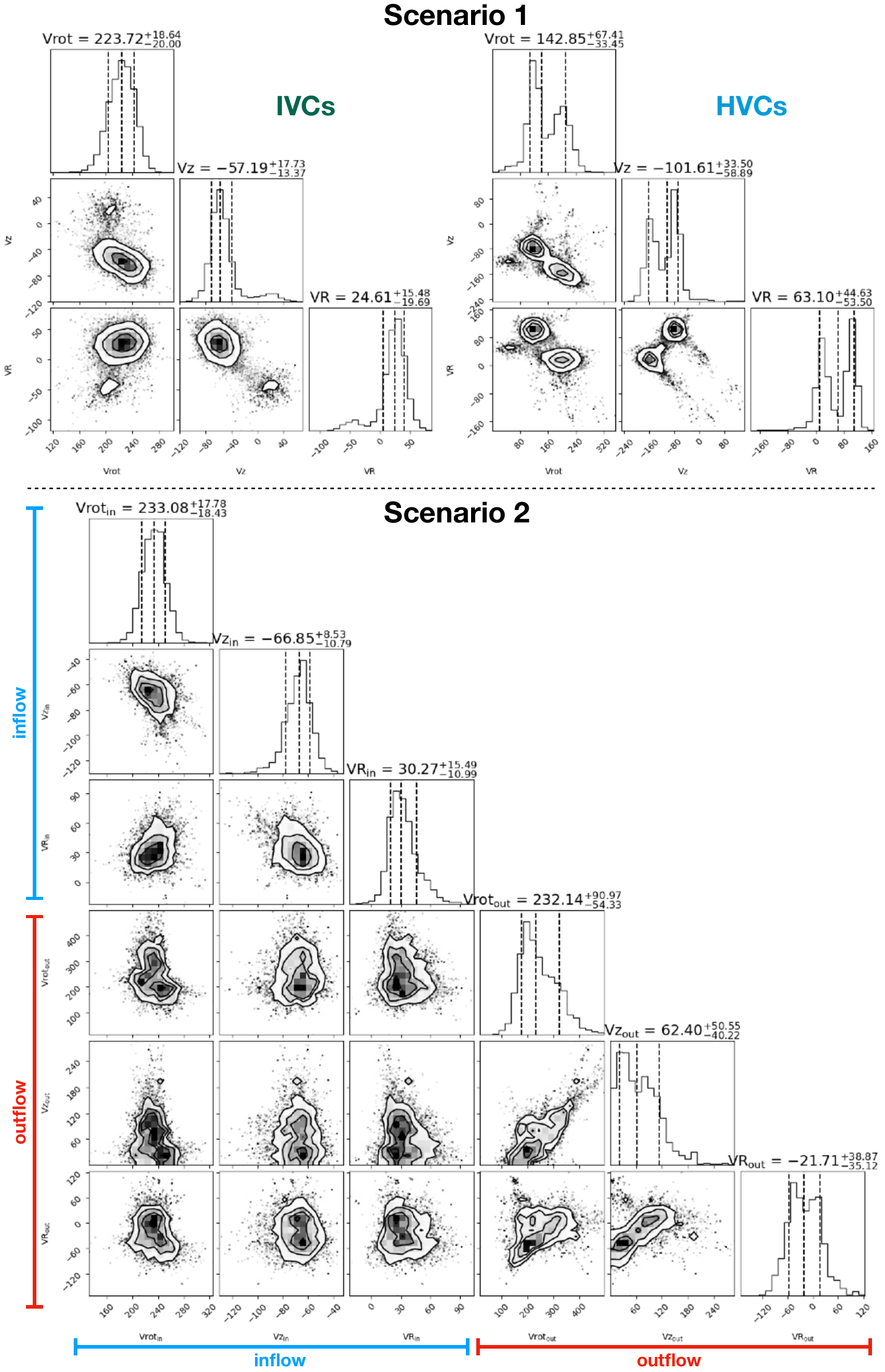}
\caption{Corner-plots showing the correlation between the various parameters used in our kinematic models of the Galactic DHI. The \emph{top} panels refer to our IVCs+HVCs scenario (Section \ref{ss-scenario1}), the \emph{bottom} panels refer to our inflow+outflow scenario (Section \ref{ss-scenario2}).}
\label{f-corner}
\end{figure*}

\clearpage 

\end{document}